\documentclass[10pt,final,journal]{IEEEtranTCOM}
\usepackage[utf8]{inputenc}
\usepackage{amsmath}
\usepackage{amsfonts}
\usepackage{cleveref}
\usepackage[final]{graphicx}
\usepackage{acronym}
\usepackage{algorithm}
\usepackage{algorithmic}
\usepackage{mathptmx}
\usepackage{color}

\usepackage{tikz}
\usetikzlibrary{calc}

\newcommand{\positiontextbox}[4][]{%
	\begin{tikzpicture}[remember picture,overlay]
		\node[inner sep=3pt, fill=yellow,align=left,draw,line width=1pt,#1] at ($(current page.north west) + (#2,-#3)$) {\parbox{.95\paperwidth}{#4}};
	\end{tikzpicture}%
}

\def\B{\mathbf}

\crefname{figure}{Fig.}{Figs.}

\newacro{SIMO}{Single-Input and Multiple-Output}
\newacro{MIMO}{Multiple-Input and Multiple-Output}
\newacro{MISO}{Multiple-Input and Single-Output}
\newacro{MAC}{Multiple Access Channel}
\newacro{JSCC}{Joint Source-Channel Coding}
\newacro{SNR}{Signal-to-Noise Ratio}
\newacro{AWGN}{Additive White Gaussian Noise}
\newacro{SQLC}{Scalar Quantizer Linear Coding}
\newacro{ML}{Maximum Likelihood}
\newacro{CSI}{Channel State Information}
\newacro{MAP}{Maximum A Posteriori}
\newacro{MMSE}{Minimum Mean Squared Error}
\newacro{MSE}{Mean Squared Error}
\newacro{pdf}{probability density function}
\newacro{OPTA}{Optimal Performance Theoretically Attainable}
\newacro{SDR}{Signal-to-Distortion Ratio}
\newacro{SOC}{Second Order Cone}
\newacro{KKT}{Karush-Kuhn-Tucker}
\newacro{SINR}{Signal-to-Interference-Noise Ratio}
\newacro{CDMA}{Code Division Multiple Access}
\newacro{MRT}{Maximum Ratio Transmitter}
\newacro{BD}{Block Diagonalization}
\newacro{BC}{Broadcast Channel}
\newacro{MRC}{Maximum Ratio Combiner}
\newacro{WSN}{Wireless Sensor Network}
\newacro{AMRT}{Aligned MRT}
\newacro{SDP}{Semidefinite Programming}
\newacro{SVD}{Singular Value Decomposition}
\normalsize

\newtheorem{theorem}{Theorem}[section]
\newtheorem{lemma}[theorem]{Lemma}

\begin{document}

	\onecolumn
	\begingroup
	
	\setlength\parindent{0pt}
	\fontsize{14}{14}\selectfont
	
	\vspace{1cm} 
	\textbf{This is an ACCEPTED VERSION of the following published document:}
	
	\vspace{1cm} 
	P. Suárez-Casal, J. P. González-Coma, O. Fresnedo and L. Castedo,``Design of Linear Precoders for Correlated Sources in MIMO Multiple Access Channels'', \textit{IEEE Transactions on Communications}, vol. 66, n.o 12, pp. 6110-6122, Aug 2018, doi: 10.1109/TCOMM.2018.2863362
	
	\vspace{1cm} 
	Link to published version: https://doi.org/10.1109/TCOMM.2018.2863362
	
	\vspace{3cm}
	
	\textbf{General rights:}
	
	\vspace{1cm} 
	\textcopyright 2018 IEEE. This version of the article has been accepted for publication, after peer review. {Personal use of this material is permitted. Permission from IEEE must be obtained for all other uses, in any current or future media, including reprinting/republishing this material for advertising or promotional purposes, creating new collective works, for resale or redistribution to servers or lists, or reuse of any copyrighted component of this work in other works.}
	\twocolumn
	\endgroup
	\clearpage

\title{Design of Linear Precoders for Correlated Sources in MIMO Multiple Access Channels}

\author{Pedro~Suárez-Casal,~Jose P. González-Coma~\IEEEmembership{Member,~IEEE},
		Óscar~Fresnedo~\IEEEmembership{Member,~IEEE},
       Luis~Castedo,~\IEEEmembership{Senior Member,~IEEE}%
\thanks{José P. González-Coma is with University of A Coruña, CITIC, Spain, e-mail: jose.gcoma@udc.es.}
\thanks{Pedro Suárez-Casal, Óscar Fresnedo and Luis Castedo are with the Department
of Computer Engineering, University of A Coruña, 15071, A Coruña, Spain,
e-mail: \{pedro.scasal, oscar.fresnedo, luis\}@udc.es.}}

\maketitle

\begin{abstract}
This work focuses on distributed linear precoding when users transmit correlated information over a fading \acl{MIMO} \acl{MAC}. Precoders are optimized in order to minimize the sum-Mean Square Error (MSE) between the source and the estimated symbols. When sources are correlated, minimizing the sum-MSE results in a non-convex optimization problem. Precoders for an arbitrary number of users and transmit and receive antennas are thus obtained via a projected steepest-descent algorithm and a low-complexity heuristic approach. For the more restrictive case of two single-antenna users, a closed-form expression for the minimum sum-MSE precoders is derived. Moreover, for the scenario with a single  receive antenna and any number of users, a solution is obtained by means of a semidefinite relaxation. Finally, we also consider precoding schemes where the precoders are decomposed into complex scalars and unit norm vectors.  
Simulation results show a significant improvement when source correlation is exploited at precoding, especially for low SNRs and when the number of receive antennas is lower than the number of transmitting nodes.

\end{abstract}

\acresetall


\IEEEpeerreviewmaketitle

\positiontextbox{11cm}{27cm}{\footnotesize \textcopyright 2018 IEEE. This version of the article has been accepted for publication, after peer review. Personal use of this material is permitted. Permission from IEEE must be obtained for all other uses, in any current or future media, including reprinting/republishing this material for advertising or promotional purposes, creating new collective works, for resale or redistribution to servers or lists, or reuse of any copyrighted component of this work in other works. Published version:
	https://doi.org/10.1109/TCOMM.2018.2863362}

\section{Introduction}
\IEEEPARstart{T}{he} simultaneous transmission of information from spatially separated devices often occurs in many wireless communications applications like \acp{WSN} and mobile cellular networks. We focus on the \ac{MAC}, also called uplink channel, where multiple user terminals transmit their data to one centralized node. We further consider the more general \ac{MIMO} case where both the receiver and the individual users are equipped with several antennas. More specifically, we address the transmission of correlated information over the fading \ac{MAC} using linear precoding.

\acp{WSN} is an example of a scenario where the assumption of sources transmitting statistically independent information does not often hold because the information is usually correlated. For example, the measurements of a parameter of interest (temperature, humidity, etc.) using sensors placed in a given area often produce correlated data. The traditional way to optimize such communication systems consists of maximizing the throughput by removing the source correlation and then protecting the relevant data with an appropriate channel coding scheme. This is known as the separation principle where source and channel coding are optimized separately \cite{shannon48}. This strategy has been shown to be optimal for multiple scenarios under certain circumstances (large block lengths, high delay, etc.), but this no longer holds for others like the transmission of correlated data over a \ac{MAC} \cite{lapidoth10,gunduz09}. 

An alternative approach consists in designing the transmission scheme to minimize the signal distortion while exploiting the source correlation. In this case, the source symbols are directly transformed into the corresponding channel symbols using appropriate encoding mappings. Some examples of this strategy in MAC communications are linear mappings \cite{lapidoth10,Lee76}, non-linear mappings \cite{Karlsson11,Mehmetoglu15,Suarez17} or schemes based on vector quantizers \cite{lapidoth10,Wernersson09B,floor15}. Linear mappings show near-optimal performance for low \ac{SNR} values, whereas non-linear approaches more efficiently exploit high correlations in the high \ac{SNR} region \cite{Wernersson09}. However, current strategies to optimize non-linear mappings for time-varying channels are computationally unaffordable in practical applications, and the alternative is to use parametric mappings adapted to the channel conditions with a small number of parameters. Linear schemes and parametric non-linear mappings are in general suboptimal for fading channels, but their performance can be improved using precoding techniques that exploit the channel knowledge at transmission. 

In multiuser scenarios assuming uncorrelated information, the channel capacity can be approached with non-linear strategies such as Costa precoding \cite{costa83} or Tomlinson-Harashima precoding \cite{Tomlinson71,Harashima72}. Linear precoding, however, often provides a reasonable performance with much lower complexity \cite{Spencer04,Zhengang04}. Linear precoders usually aim at optimizing metrics like the \ac{SINR} \cite{Schubert04}, the sum-\ac{MSE} \cite{hunger08,Tenenbaum04}, or the balancing of individual distortions \cite{shi08,GoGrJoCa17}. In general, these approaches improve the performance of digital communications and, also, lowering the \ac{MSE} is optimal from the point of view of analog communications where the ultimate goal is minimizing the signal distortion. 

In this work, we address the optimization of distributed linear precoding techniques that exploit the source correlation to minimize the signal distortion in the fading \ac{MIMO} \ac{MAC}. Unlike the uncorrelated sources case, the source covariance matrix is no longer diagonal and the resulting optimization problems cannot be reformulated in a convex form. In spite of that, optimal solutions are obtained for some particular scenarios, while suboptimal approaches are considered in the more general case. The optimization of linear transceivers according to the \ac{MSE} metric has already been considered for correlated sources and single-user \ac{MIMO} channels \cite{Lee76}, and for the case of bivariate Gaussian sources and collaborative Gaussian \ac{MAC} \cite{dabeer07}. The design of linear filters based on the \ac{SINR} is addressed in \cite{Yang95}. However, they considered an array model and a particular error model to perform the analysis, and the precoding scalar factors are set to one. Also, some authors addressed the linear precoding of correlated sources through the design of \ac{CDMA} signatures where the users transmit their symbols using several channel uses over a fading \ac{MAC} and the receiver has a single antenna \cite{Acharya04,Yahampath15}. 

The remaining of this paper is organized as follows. \Cref{sec:system} presents the communication model considered in this work. In \Cref{sec:precoder_sumMSE}, we address the design of linear precoding schemes for the general \ac{MIMO} \ac{MAC} and present three approaches which provide different alternatives to exploit source correlation.
\Cref{sec:SDR} focuses on the \ac{MISO} \ac{MAC} scenario where the linear precoders are obtained by using \ac{SDP} to simplify the optimization problem.  
\Cref{sec:TwoUsers} provides an analytical solution for the two-user \ac{SIMO} \ac{MAC} precoder. \Cref{sec:results} presents the results of simulation experiments carried out to illustrate the performance of the proposed strategies. Finally, \Cref{sec:conclusions} is devoted to the conclusions.

\subsection{Contributions}
In this work, we propose different approaches to optimize the linear precoding of correlated Gaussian sources according to sum-\ac{MSE}:
\begin{itemize}

	\item A closed-form expression for the optimal linear precoder that minimizes the sum-\ac{MSE} in the two-user \ac{SIMO} \ac{MAC} scenario.
	
	\item An \ac{SDP} approach for scenarios with an arbitrary number of users and a single receive antenna. This solution leads to a convex optimization problem that can be efficiently solved by interior point methods.
		
	\item For the general case of multiple antennas at transmission and reception, two approaches are considered. The first one is a heuristic approximation assuming all users are decoded with the same receive filter. The second one is based on splitting the precoding vectors into a complex-valued scalar and a unit-norm direction vector. Direction vectors are obtained using either \ac{MRT} or nullspace-directed SVD (Nu-SVD) that do not take into account source correlation. However, gain factors are determined to minimize the sum-MSE by exploiting source correlation.
\end{itemize}

\section{System Model} \label{sec:system}

\begin{figure}
	\centering
	\includegraphics[scale=0.4]{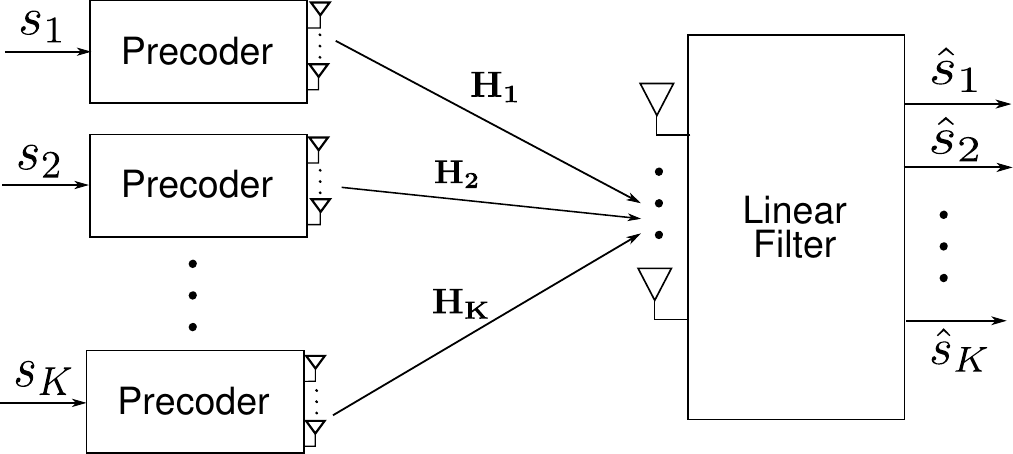}
	\caption{Block-diagram of a fading \ac{MIMO} \ac{MAC} system with $K$ users, linear precoders, and a linear receiver.}
	\label{fig:system_model}
\end{figure}
Let us consider the fading \ac{MIMO} \ac{MAC} system model shown in \Cref{fig:system_model} where $K$ individual users with $N_T$ antennas each\footnote{For the sake of simplicity, we assume all users have the same number of transmit antennas. The extension of the ideas in this work to \ac{MIMO} \acp{MAC} with different number of transmit antennas per user is straightforward.} send correlated information to a common receiver with $N_R$ antennas. The $k$-th user sends one source symbol per channel use $s_k$. The vector $\mathbf{s} = [s_1, s_2, \ldots, s_K]^T$ comprising all users source symbols follows a multivariate complex-valued distribution with zero mean and covariance matrix $\mathbf{C}_{\mathbf{s}}= E\left[\mathbf{s}\mathbf{s}^H\right]$. The superindices $^T$ and $^H$ represent the transpose and Hermitian operators, respectively. The elements $[\mathbf{C}_{\mathbf{s}}]_{i,j} = \rho_{i,j}$ represent the correlation between the $i$-th and $j$-th source symbols. Without loss of generality, we will assume that $\rho_{i,i} = 1~ \forall i$ and $0\leq\rho_{i,j} \leq 1 ~\forall i, j$ such that $i \ne j$.

The source symbols are individually precoded at each user using a linear scheme and the resulting symbols are then transmitted to the receiver over their corresponding fading channels. We will consider a small-scale fading model, disregarding the attenuation caused by free space propagation and other effects.
The received MAC signal is represented as
\begin{equation}
\mathbf{y} = \sum_{k=1}^K \B{H}_k\B{p}_k s_k+ \mathbf{n},
\label{eq:simo_mac_model}
\end{equation}
where $\B{p}_k\in \mathbb{C}^{N_T\times 1}$ and $\mathbf{H}_k\in \mathbb{C}^{N_R\times N_T}$ are the $k$-th user linear precoding vector and \ac{MIMO} channel response, respectively; and  $\mathbf{n} = [n_1, \ldots, n_{N_R}]^T$ represents the complex-valued \ac{AWGN} such that $\mathbf{n} \sim \mathcal{N}_{\mathbb{C}}(\mathbf{0},\sigma_n^2\mathbf{I})$. An individual power constraint is imposed at each transmitter, which is given by $\B{p}_k^H\B{p}_k \le T_k, k=1,\ldots,K$ since the power of the source symbols is assumed to be equal to $1$. The received signal \eqref{eq:simo_mac_model} can be rewritten in a more compact form as 
\begin{align}
\B{y} = \B{H}\B{P}\B{s} +\B{n},
\end{align}
where $\B{H} = [\B{H}_1 \cdots \B{H}_K]$ and $\B{P}=\operatorname{blockdiag}\left(\B{p}_1,\ldots,\B{p}_K\right)$, with $\operatorname{blockdiag}(\cdot)$ the operator that constructs a block-diagonal matrix and puts the matrices received as arguments on its main diagonal, setting the off-diagonal elements to zero. Finally, a linear estimate of the source symbols $\hat{\B{s}} = [\hat{s}_1,\hat{s}_2,\ldots, \hat{s}_K]^T = \B{W}^H\B{y}$ is computed, $\B{W}$ being the linear receiver filter response.

In the ensuing sections, we consider several strategies to determine the linear precoders in $\B{P}$ for the described \ac{MIMO} \ac{MAC} system model.

\section{\ac{MIMO} \ac{MAC} Linear Precoding}\label{sec:precoder_sumMSE}

Along this work, the distortion between the source and the estimated symbols will be measured in terms of the sum-\ac{MSE} criterion as
\begin{align}
\xi_{\text{sum}} = \mathbb{E}\left[\sum\limits_{k=1}^K |s_k - \hat{s}_k|^2\right] = \mathbb{E}\left[\operatorname{tr}\left((\B{s} - \hat{\B{s}})(\B{s} - \hat{\B{s}})^H\right)\right].
\label{eq:sum-MSE}
\end{align}
For given precoders and channel responses, $\B{P}$ and $\B{H}$, the linear \ac{MMSE} receiving filter is
\begin{align}
\B{W}^H = \B{C}_{\B{s}}\B{P}^H\B{H}^H\left(\B{H}\B{P}\B{C}_{\B{s}}\B{P}^H\B{H}^H + \frac1{\sigma_n^2}\B{I}\right)^{-1}.
\label{eq:MMSEestim}
\end{align}
Substituting \eqref{eq:MMSEestim} into \eqref{eq:sum-MSE} leads to the following expression for the sum-\ac{MSE} \cite{FundamentalsSSP}
\begin{align}
\xi_{\text{sum}} =  f(\B{P}) = \operatorname{tr}\left( \frac1{\sigma_n^2}\B{P}^H\B{H}^H\B{H}\B{P} + \B{C}_s^{-1} \right)^{-1}.
\label{xisum}
\end{align}
The optimal linear \ac{MIMO} \ac{MAC} precoder that minimizes the sum-MSE is hence determined from the following constrained optimization problem 
\begin{align}
\underset{\B{p}_1,\ldots,\B{p}_K}{\arg \min}~~&  \operatorname{tr}\left( \frac1{\sigma_n^2}\B{P}^H\B{H}^H\B{H}\B{P} + \B{C}_s^{-1} \right)^{-1}\label{eq:opt_problem}\\
s.t. ~~& \B{p}_k^H\B{p}_k\le T_k, \forall k\in[1,K],\notag\\
~~ & \B{P} = \operatorname{blockdiag}\left(\B{p}_1,\ldots,\B{p}_K\right),
\notag
\end{align}
where $\B{p}_k^H\B{p}_k\le T_k$ represents the individual power constraints and $\B{P} = \operatorname{blockdiag}\left(\B{p}_1,\ldots,\B{p}_K\right)$ arises from the restriction that users do not cooperate and, therefore, the source symbols are individually precoded at each transmitter.

The optimization problem in \eqref{eq:opt_problem} is non-convex on $\B{P}$ when sources are correlated. When $\B{C_s}=\B{I}$ (i.e. uncorrelated sources), \eqref{eq:opt_problem} can be reformulated in convex form over the transmit covariance matrices $\B{P}\B{C_s}\B{P}^H$ using the matrix inversion lemma (see \cite{hunger08} and references therein). A convex reformulation has also been adopted when designing CDMA signatures for the transmission of correlated sources with one receive antenna \cite{Acharya04,Yahampath15}. By dropping the restrictions over the shape of $\B{P}$, e.g., allowing users to cooperate in the \ac{MAC} or considering a \ac{BC}, idempotent matrices $\B{C_s}$ are also valid. %
Nevertheless, none of these strategies are valid to obtain a convex reformulation of \eqref{eq:opt_problem},
since they cannot be applied to the case of distributed non-cooperative users as is usually the case in the \ac{MAC}. 
However, some interesting conclusions can be drawn for asymptotic cases assuming that the entries of the channel matrix are independent and identically distributed (i.i.d.) complex-valued Gaussian random variables with zero mean and unit variance, i.e., $[\B{H}]_{i,j}\sim \mathcal{N}_{\mathbb{C}}(0, 1)$:
\begin{itemize}
\item When $N_T=1, N_R\rightarrow \infty$ and the number of users is fixed, the product $\frac{1}{N_R}\B{H}^H\B{H}$ converges almost surely to the identity matrix \cite{Ma10}. In this case, the MSE can be approximated as
\begin{align}\operatorname{tr}&\left( \frac1{\sigma_n^2}\B{P}^H\B{H}^H\B{H}\B{P} + \B{C}_s^{-1}\right)^{-1} \\
&= \frac{1}{N_R}\operatorname{tr}\left( \frac1{N_R\sigma_n^2}\B{P}^H\B{H}^H\B{H}\B{P} + \frac{1}{N_R}\B{C}_s^{-1}\right)^{-1}\notag\\
&\approx \frac{1}{N_R}\operatorname{tr}\left( \frac1{\sigma_n^2}\B{P}^H\B{P}\right)^{-1} = \frac{\sigma_n^2}{N_R}\sum_{k=1}^K \frac{1}{ |p_k|^2}\notag,
\end{align}
and therefore the optimum power allocation is $p_k^o = \sqrt{T_k}, \forall k$.
\item In the high SNR regime ($\sigma_n^2 \rightarrow 0$), the weight of the source correlation matrix in \eqref{eq:opt_problem} is negligible. Therefore, the solutions for uncorrelated scenarios apply and $p_k^o = \sqrt{T_k}$.
\end{itemize}
In the described scenarios, good solutions for the optimal linear precoder are expected to be obtained without considering source correlation. In the next subsections, different approaches are studied for the design of linear precoders that exploit the source correlation.
\vspace{-0.2cm}
\subsection{Projected Gradient Algorithm}
\label{sec:Gradient}
A first approach to solve \eqref{eq:opt_problem} is by means of a projected steepest-descent or gradient algorithm, where the precoding matrix $\B{P}$ is updated at each iteration by
\begin{equation}
\B{P}^{(i+1)} = t\left( \B{P}^{i} + \mu^{i} \nabla f(\B{P}^i) \right),
\label{eq:precoder_gradient}
\end{equation}
where $i$ indicates the iteration number, $\mu$ is the algorithm step size, $\nabla f(\B{P})$ is the gradient vector of the cost function $\xi_{\text{sum}} =  f(\B{P})$ with respect to $\B{P}$ given by
\begin{align*}
\nabla f(\B{P}) = \frac{\partial f(\B{P})}{\partial \B{P}^*} = -\frac{2}{\sigma_n^2}\B{H}^H\B{H}\B{P}\left( \frac1{\sigma_n^2}\B{P}^H\B{H}^H\B{H}\B{P} + \B{C}_s^{-1}\right)^{-2},
\end{align*}
and $t\left(\cdot\right)$ projects the precoder into the space of feasible solutions given by the set of constraints. In our case, the projection normalizes the power of the user precoders in $\B{P}$ that exceeds their corresponding constraints and sets the off-diagonal elements to zero. Hence, $t\left(\cdot\right)$ is a projector onto a closed convex set, thus it is non-expansive,
i.e., $\|t(\B{P}_1)-t(\B{P}_2)\|_F\leq \|\B{P}_1-\B{P}_2\|_F$ \cite[Prop. 2.1.3]{Bertsekas76}.
Applying this property and with $\B{P}^o$ the optimum of the cost function, we get $\|\B{P}^{(i)}-\B{P}^o\|_{F}^2 \ge \|\B{P}^{(i+1)}-\B{P}^o\|_{F}^2 \ge \|t(\B{P}^{(i+1)})-\B{P}^o\|_{F}^2$ and convergence to a local optimum is guaranteed as long as the step size is properly adjusted at each iteration by using, e.g., a line or Armijo's search \cite{Bertsekas76}. Since the problem is non-convex, the convergence of algorithm \eqref{eq:precoder_gradient} to the global optimal solution is not guaranteed, but it always achieves a stationary point.

The computational complexity of this algorithm can be significant since the number of steps to achieve a solution with an error below some given threshold is unbounded in general. In practical terms, we have checked that its performance is as good as any other method we have investigated, and hence we will use it as a reference for comparison.
\subsection{Aligned \ac{MRT}}
\label{sec:amrt}
In this section, we present a low complexity approach based on aligning the different users to a single receive direction. This simple design is inspired by the idea that only a small part of the signals from other users is actually an interference when the source symbols are highly correlated, and hence the received symbols can be combined constructively. This approach is especially suitable for low SNRs since the impact of the source correlation on the sum-\ac{MSE} minimization is more significant at the low SNR regime, while it vanishes for high \acp{SNR}. 

The selection of the receive direction is of key importance since it will determine the overall system performance. Unfortunately, the selection of the best direction taking into account the channels for all the users is a combinatorial problem. Hence, we propose a heuristic where the precoders are designed assuming that the receive filter is common for all users, i.e.
\begin{equation}
	\underset{\B{g},\B{p}}{\arg \max}\max_k|\B{g}^H\B{H}_k\B{p}|, ~~
s.t.~~\Vert \B{g} \Vert = 1, \Vert \B{p} \Vert = 1,
\end{equation}
where the receive direction $\B{g}$ is chosen as the channel left singular vector corresponding to the maximum singular value among all users. %
Given the direction vector $\B{g}$, the precoder for the $k$-th user is computed as the vector $\B{p}_k$ that maximizes the product $|\B{g}^H\B{H}_k\B{p}_k|$, i.e., 
\begin{equation}
	\B{p}_k=\frac{\sqrt{T_k}}{\|\B{H}_k^H\B{g}\|}\B{H}_k^H\B{g}.
	\label{eq:alignedPrec}
\end{equation}
This approach hence maximizes the power of the correlated signal in the receive direction $\B{g}$, because the users encode their symbols to allocate the available power into a single spatial direction.
Finally, for the precoders \eqref{eq:alignedPrec}, the receive filters are computed according to \eqref{eq:MMSEestim}.  This strategy will be termed \ac{AMRT}.
 
\subsection{\ac{MRT} and Nu-SVD for Correlated Sources}
\label{subsec:MIMOPrecoding}

In this section, we focus on \acl{MRT} (MRT) and  nullspace-directed \ac{SVD} algorithm (Nu-SVD), two representative linear precoding strategies for the \ac{MIMO} \ac{BC} and \ac{MAC}. We next explain how to integrate \ac{MRT} and Nu-SVD precoding in a \ac{MIMO} \ac{MAC} signaling scheme that exploits source correlation. The key step is to decompose the individual precoders as $\B{p}_k = \gamma_k \B{u}_k$, where $\gamma_k \in \mathbb{C}$ is a complex-valued gain factor and $\B{u}_k \in \mathbb{C}^{N_T\times 1}$ is a unit-norm direction vector, i.e., $\|\B{u}_k\|^2 = 1$. When considering \ac{MRT}, the directions $\B{u}_k$ are chosen to maximize the receiver SNR, i.e. $\B{u}_k^{\text{MRT}} = \B{v}_{k,1}$, where $\B{v}_{k,1}$ is the right singular vector with the largest singular value in the \ac{SVD} $\B{H}_k = \B{U}_k\boldsymbol{\Sigma}_k\B{V}_k^H$ \cite{BjBeOt14}.

An alternative strategy is the iterative nullspace-directed SVD algorithm proposed for precoding in the \ac{MIMO} \ac{BC} \cite{Zhengang04} \cite[Section V]{Spencer04}. It can be applied to the \ac{MIMO} \ac{MAC} by invoking the \ac{MSE}-duality to transform the receive filters in the \ac{BC} into the precoders in the \ac{MAC} \cite{hunger08,bogale12}. In general, this approach iteratively searches for matrices  $\B{P}_{\text{NSVD}} = \operatorname{blockdiag}\left(\B{u}_1^{\text{NSVD}}, \ldots, \B{u}_K^{\text{NSVD}}\right)$ and $\B{W}_{\text{NSVD}}$ such that the inter-user interference is canceled and it hence provides solutions that satisfy the condition $\B{W}_{\text{NSVD}}^H\B{H}\B{P}_{\text{NSVD}} = \B{I}$. This property holds both  in the \ac{BC} and the \ac{MAC}.

In uncorrelated sources scenarios, the solutions for the gain factors $\gamma_k$ are trivial and are set to the maximum available power, i.e. $\gamma_k = \sqrt{T_k}$.  However, source correlation can be exploited at transmission if the gain factors $\gamma_k$ are optimized to minimize the sum-MSE, as shown analytically in \Cref{sec:TwoUsers} for the two-user \ac{SIMO} \ac{MAC} and numerically in \Cref{sec:results} for a larger number of users. This can be done by applying the projected steepest-descent algorithm in \eqref{eq:precoder_gradient} to the equivalent channels $\tilde{\B{h}}_k = \B{H}_k\B{u}_k$. This approach reduces the dimension of the search space since the original problem with $N_TK$ variables is transformed into a search over the $K$ gain coefficients. %

\section{Multiuser SISO scenario}
\label{sec:SDR}

For the particular scenario with a single receive antenna and a single transmit antenna per user, i.e., $N_R=1, N_T=1$, the channel matrix reduces to a row vector $\B{h}\in\mathbb{C}^{1\times N_TK}$ and the sum-\ac{MSE} expression \eqref{xisum} can be rewritten by applying the matrix inversion lemma as
\begin{align}
\xi_{\text{sum}} =& \operatorname{tr}\left( \frac1{\sigma_n^2}\B{P}^H\B{h}^H\B{h}\B{P} + \B{C}_s^{-1} \right)^{-1} \notag\\
= & K - \operatorname{tr} \left( \B{C}_s\B{P}^H\B{h}^H \left(\B{h}\B{P}\B{C}_s\B{P}^H\B{h}^H   + \sigma_n^2\right)^{-1} \B{h}\B{P}\B{C}_s\right) \notag\\ 
 =& K - \frac {\B{p}^H\B{H} \B{C}_s^2\B{H}^H\B{p} }{\B{p}^H\B{H}\B{C}_s\B{H}^H\B{p} + \sigma_n^2},
\end{align}
where $\B{H} = \operatorname{diag}(\B{h})$, and $\B{p}\in \mathbb{C}^{N_TK\times 1}$ stacks the diagonal elements of $\B{P}^H$. Hence, the optimization problem can be restated as
\begin{align}
\underset{\B{p}}{\arg \max}~~&  \frac {\B{p}^H\B{H} \B{C}_s^2\B{H}^H\B{p} }{\B{p}^H\B{H}\B{C}_s\B{H}^H\B{p}   + \sigma_n^2}\label{eq:problFormQuotient}\\
s.t. ~~& \B{p}^H \B{e}_k\B{e}_k^T \B{p}\le T_k, \forall k\in[1,K],\notag
\end{align}
where $\B{e}_k$ denotes the indicator vector. Following a similar approach to \cite{beck2010minimizing}, the system is homogenized with $\B{p} = \B{q}/t$, and rewritten as
\begin{align}
\underset{\B{q},t}{\arg \max}~~&  f(\B{q},t)=\frac {\B{q}^H\B{H} \B{C}_s^2\B{H}^H\B{q} }{\B{q}^H\B{H}\B{C}_s\B{H}^H\B{q}   + \sigma_n^2t^2}\\
s.t. ~~& \B{p}^H \B{e}_k\B{e}_k^T \B{p}\le t^2T_k, \forall k\in[1,K]\notag.
\end{align}
If we denote $g(\B{z}) = f(\B{q},t)$ with $\B{z}=[\B{q}^T, t]^T$, it can be seen that it is invariant to scaling, i.e. $g(k\B{z}) = g(\B{z}), k\in\mathbb{R}^+$. Hence, we can fix the denominator to a constant value to transform the problem as
\begin{align}
\underset{\B{z}}{\arg \max}~~& \B{z}^H \B{A} \B{z}\label{eq:optEquivalent}\\
s.t. ~~& \B{z}^H\B{B}\B{z} = 1,\label{eq:radius1}\\
&\B{z}^H \B{D}_k \B{z}\le 0, \forall k\in[1,K],\label{eq:constraints}\\
& t\neq 0 \label{eq:t0},
\end{align}
where $\B{A} = \operatorname{blockdiag}(\B{H}\B{C}_s^2\B{H}^H, 0)$, $\B{B} = \operatorname{blockdiag}(\B{H}\B{C}_s\B{H}^H, \sigma_n^2)$, and $\B{D}_k =\operatorname{blockdiag}(\B{e}_k\B{e}_k^T, -T_k)$, which is a non-convex Quadratically Constrained Quadratic Programming (QCQP) problem. This problem simplifies by dropping condition \eqref{eq:t0} because for $t=0$ the constraints in \eqref{eq:constraints} are fulfilled only with $\B{q}=\B{0}$, but then \eqref{eq:radius1} is not satisfied. Hence, the feasible set is empty for $t=0$. We now define $\B{Z} = \B{z}\B{z}^H$ and rewrite the optimization problem \eqref{eq:optEquivalent}-\eqref{eq:t0} as
\begin{align}
\underset{\B{Z}}{\arg \max}~~& \operatorname{tr}(\B{A} \B{Z})\label{eq:SDR}\\
s.t. ~~& \operatorname{tr}(\B{B} \B{Z}) = 1, \notag\\
&\operatorname{tr}(\B{D}_k \B{Z})\le 0, \forall k\in[1,K]\notag,\\
&\B{Z}\succeq 0\notag,
\end{align}
where a constraint $\operatorname{rank}(\B{Z})=1$ has been dropped. The problem in \eqref{eq:SDR} is a \ac{SDP} that can be efficiently solved with interior point methods, and it is a convex approximation of the QPQC in \eqref{eq:optEquivalent}-\eqref{eq:t0}. Their solutions are not guaranteed to coincide because it is possible, for an optimal $\B{Z}^o$ in \eqref{eq:SDR}, that $\operatorname{rank}(\B{Z}^*)>1$. In that case, it is necessary to resort to techniques that obtain rank-1 approximations of $\B{Z}^o$ to estimate the optimal solution of \eqref{eq:optEquivalent}-\eqref{eq:t0}. In our problem, we can obtain optimal rank-1 approximations when the solutions to \eqref{eq:SDR} have a particular structure using the following lemma.
\begin{lemma}
Given an optimal solution $\B{Z}^o$ for a problem in the form of \eqref{eq:SDR} such that
\begin{align}
\B{Z}^o = \left(
\begin{array}{cc}
\bar{\B{Z}} & \B{v}\\
\B{v}^H & w
\end{array}
\right),
\end{align}
then, if $\operatorname{rank}(\bar{\B{Z}}) = 1$ with $\bar{\B{Z}}=\B{u}\B{u}^H$, an optimal rank-1 approximation to $\B{Z}^o$ is 
\begin{align}
\B{Z}^+ = \left(
\begin{array}{cc}
\B{u}\B{u}^H & \B{u}\sqrt{w}\\
\B{u}^H\sqrt{w} & w
\end{array}
\right),
\end{align}
and the optimal solution to problem \eqref{eq:optEquivalent} is $\B{q}^o = \B{u}$ and $t^o=\sqrt{w}$.
\begin{proof}
	This is straightforward by checking the structure of $\B{A}$, $\B{B}$ and $\B{D}_k$, where the vectors $\B{v}$ from the optimal solution are multiplied by zeros. Hence, since $\operatorname{tr}(\B{AZ}^o) = \operatorname{tr}(\B{AZ}^+)$,  and the restrictions are fulfilled, i.e., $\operatorname{tr}(\B{B} \B{Z}^+) = 1$ and $\operatorname{tr}(\B{D}_k \B{Z}^+)\le 0, \forall k\in[1,K]$, we conclude that $\B{Z}^+$ is also optimal. Hence, it optimizes \eqref{eq:optEquivalent} with the solutions given by the lemma.
\end{proof}
\end{lemma}

In case $\operatorname{rank}(\bar{\B{Z}})>1$, we choose the eigenvector with the largest associated eigenvalue as approximation. Finally, the precoder is defined as $\B{p}^o = \B{u}/\sqrt{w}$.
We have experimentally checked that, if $N_T=1$, the solutions for $\bar{\B{Z}}$ obtained through the SDP are always rank-1. Hence, they are always optimal. This approach can be naturally extended to more than one transmit antenna per user by modifying the constraints, but we found that the solutions of the relaxed version of the problem are not always rank-1. Hence, we restrict our results in \Cref{sec:results} to $N_T=1$.

\section{Optimal Linear Precoding for the Two-User \ac{SIMO} \ac{MAC}} \label{sec:TwoUsers}

The sum-MSE minimization problem \eqref{eq:opt_problem} is difficult to solve analytically for the general \ac{MIMO} \ac{MAC}. However, in this section, we will show that it can be analytically solved for the more specific two-user \ac{SIMO} \ac{MAC}, i.e., for the case of two single-antenna users and an arbitrary number of receive antennas. In this scenario, the precoder matrix reduces to
\begin{align}
\B{P}=\left(
\begin{array}{cc}
\sqrt{P_1}e^{-j\pi\phi_1} & 0\\
0 & \sqrt{P_2}e^{-j\pi\phi_2}
\end{array} 
\right),
\end{align}
where $P_1$ and $P_2$ represent the power allocated to the single-antenna users $1$ and $2$, respectively, while $\phi_1$ and $\phi_2$ represent their phase shifts. The source covariance matrix also reduces to
\begin{align}
\mathbf{C}_{\mathbf{s}}=\left(
\begin{array}{cc}
1 & \rho\\
\rho  & 1
\end{array} 
\right),
\end{align}
where $\rho = E[s_1 s_2^*]$ represents the source symbols correlation between the two users. Without loss of generality, we assume $\rho$ is real-valued. Particularizing \eqref{eq:sum-MSE} for this scenario, the sum-MSE is given by \cite{lapidoth10}
\begin{align}
\xi_{\text{sum}}(P_1,P_2,\phi_d,\sigma_n^2,\rho,\mathbf{H}) =  &\sigma_n^2\frac{2\sigma_n^2 + (1- \rho^2) (P_1\Vert \B{h}_1\Vert^2 + P_2\Vert \B{h}_2\Vert^2)}
{\sigma_n^4 + \sigma_n^2\upsilon + \omega},
\label{eq:mse_2users}
\end{align}
where
\begin{align}
\upsilon = &P_1\Vert \B{h}_1\Vert^2 
+ P_2\Vert \B{h}_2\Vert^2
+ 2\rho\sqrt{P_1P_2}\Re\{ e^{-j\pi\phi_d}\B{h}_1^H\B{h}_2\},\notag\\
\omega =&P_1P_2(1-\rho^2) \left(\Vert \B{h}_1 \Vert^2\Vert \B{h}_2 \Vert^2
- | \B{h}_1^H\B{h}_2 |^2 \right) = P_1P_2(1-\rho^2) |\B{H}^H\B{H}|\notag,
\end{align}
and $\phi_d = \phi_1 - \phi_2$. The terms $\B{h}_1$ and $\B{h}_2$ correspond to the channel responses for the first and second transmitter, respectively, such that $\B{H} = [\B{h}_1 \B{h}_2]$. For two users, the optimization problem \eqref{eq:opt_problem} simplifies to
\begin{align}
\underset{P_1, P_2, \phi_d}{\arg\min} ~~~ \xi_{\text{sum}},~~ s.t. ~~~ P_1 \le T_1,\, P_2 \le T_2.
\end{align}
The optimal linear precoder is obtained with the help of the following two lemmas. The first one determines the optimal phases while the second determines the optimal power allocations for the optimal phases.

\begin{lemma}
\label{lemma:opt_phases}
The optimal phases $\phi_1^{\text{opt}}$ and $\phi_2^{\text{opt}}$ for the two-user \ac{SIMO} \ac{MAC} linear precoder must satisfy
\begin{align}
\phi_1^{\text{opt}}-\phi_2^{\text{opt}} = \phi_d^{\text{opt}} = \underset{\phi_d}{\arg\max} ~\Re\{e^{-j\pi\phi_d} \B{h}_1^H\B{h}_2\} = \arg(\B{h}_1^H\B{h}_2).
\label{eq:rotator_precoders}
\end{align}

\begin{proof}
The variable $\phi_d$ is only present in the denominator of \eqref{eq:mse_2users}, in the term  $\Re\{ e^{-j\pi\phi_d}\B{h}_1^H\B{h}_2\}$. Thus, for any transmit powers $P_1$ and $P_2$, the sum-\ac{MSE} lowers when this terms increases.
It is straightforward to see that this term is maximum for $\phi_d^{\text{opt}} = \arg(\B{h}_1^H\B{h}_2)$ because in that case $\Re\{ e^{-j\pi\phi_d}\B{h}_1^H\B{h}_2\} = |\B{h}_1^H\B{h}_2|$.
\end{proof}
\end{lemma}

Using this lemma, without loss of generality, we can assume that $\phi_1^{\text{opt}} = 0$ and therefore $\phi_2^{\text{opt}} = \phi_d^{\text{opt}}$.
Since the term $\Re\{ e^{-j\pi\phi_d}\B{h}_1^H\B{h}_2\}$ is multiplied by the correlation factor $\rho$, this optimization of the precoder phases only improves the performance in the case of correlated sources with $\rho>0$. Besides, the gain is more remarkable as the source correlation increases, as we will show in the results section.

The next step is to find the optimal power allocation that minimizes the sum-MSE assuming the optimal value for the precoder phases. Replacing $\phi_d$ by its optimal value in \eqref{eq:mse_2users} produces
\begin{align}
	\bar{\xi}_{\text{sum}}(P_1,P_2,\sigma_n^2,\rho,\mathbf{H}) =  &\sigma_n^2\frac{2\sigma_n^2 + (1- \rho^2) (P_1\Vert \B{h}_1\Vert^2 + P_2\Vert \B{h}_2\Vert^2)}
	{\sigma_n^4 + \sigma_n^2\bar{\upsilon} + \omega},
\end{align}
where
$\bar{\upsilon}=P_1\Vert \B{h}_1\Vert^2 
+ P_2\Vert \B{h}_2\Vert^2
+ 2\rho\sqrt{P_1P_2}|\B{h}_1^H\B{h}_2|.$
The following Lemma provides the optimal power allocation for given $\phi_d$.
\begin{lemma}
\label{theorem:power_alloc}
The solution to the optimal power allocation problem
\begin{align}
\underset{P_1, P_2}{\arg\min} ~~~ \bar{\xi}_{\text{sum}},~~~~ s.t.~~~ P_1 \le T_1, P_2 \le T_2
\label{eq:opt_problem_2users}
\end{align}
is
\begin{itemize}
\item  If $\rho=0$, $\rho=1$ or $|\B{h}^H_1\B{h}_2|=0$, then $P_1 = T_1$ and $P_2 = T_2$. 
\item In any other case, $P_1 = \min\left(T_1, f_1\left(T_2\right)^2\right)$ and $P_2 = \min\left(T_2, f_2\left(T_1\right)^2\right)$ where
\begin{align}
f_1(P_2, \sigma_n^2, \rho, \mathbf{H}) &=Z +
\sqrt{Z^2 + \frac{\Vert\B{h}_2\Vert^2}{\Vert\B{h}_1\Vert^2}P_2 +
	\frac{2\sigma_n^2}{(1-\rho^2)\Vert\B{h}_1\Vert^2}},\\
f_2(P_1,\sigma_n^2, \rho, \mathbf{H}) &= f_1(P_1,\sigma_n^2, \rho, [\mathbf{h}_2, \mathbf{h}_1]),
\label{eq:power_node2}
\end{align}
with
\begin{align}
Z =& \frac{ \sigma_n^2}{(1-\rho^2)|\B{h}_1^H\B{h}_2|\sqrt{P_2}} \notag
\\&+ \frac{ \sigma_n^4\Vert \B{h}_1 \Vert^2  + \left(2\sigma_n^2 + (1-\rho^2)\Vert\B{h}_2\Vert^2P_2 \right)|\B{H}^H\B{H}|P_2}{2\sigma_n^2\rho|\B{h}_1^H\B{h}_2|\Vert\B{h}_1\Vert^2\sqrt{P_2}}.\notag
\end{align}
\end{itemize}

\begin{proof}
The Lagrangian function associated to \eqref{eq:opt_problem_2users} is
\begin{align}
\mathcal{L} =  \bar{\xi}_{\text{sum}}(P_1,P_2,\sigma_n^2,\rho,\mathbf{H}) + \lambda_1(P_1 - T_1) + \lambda_2(P_2 - T_2).
\end{align}
The optimal precoder must hence satisfy the following necessary \ac{KKT} conditions \cite{karush39,kuhn51}
\begin{align}
\frac{\partial \mathcal{L}}{\partial P_1} = \sigma_n^2\frac{\alpha_1(P_1, P_2, \sigma_n^2, \rho, \mathbf{H})}{\beta(P_1, P_2, \sigma_n^2, \rho, \mathbf{H})}& + \lambda_1 = 0\label{eq:KKT1}\\
\frac{\partial \mathcal{L}}{\partial P_2} = \sigma_n^2\frac{\alpha_2(P_1, P_2, \sigma_n^2, \rho, \mathbf{H})}{\beta(P_1, P_2, \sigma_n^2, \rho, \mathbf{H})}& + \lambda_2 = 0\label{eq:KKT2}\\
\lambda_1(P_1-T_1)& = 0\label{eq:KKT3}\\
\lambda_2(P_2-T_2)& = 0,\label{eq:KKT4}
\end{align}
with
\begin{align}
\label{eq:diff_P1}
\alpha_1(P_1,& P_2, \sigma_n^2, \rho, \mathbf{H}) = -\sigma_n^4(1+\rho^2) \Vert \B{h}_1 \Vert^2  \notag\\
&- 2\sigma_n^2(1-\rho^2)|\B{H}^H\B{H}|P_2 - (1-\rho^2)^2\Vert\B{h}_2\Vert^2|\B{H}^H\B{H}|P_2^2 \notag\\
& + \sigma_n^2\rho(1-\rho^2)|\B{h}_1^H\B{h}_2| \left(\Vert \B{h}_1\Vert^2 P_1 - \Vert \B{h}_2\Vert^2 P_2\right)\sqrt{\frac{P_2}{P_1}} \notag\\
& - 2\sigma_n^4\rho |\B{h}_1^H\B{h}_2|\sqrt{\frac{P_2}{P_1}},\\
\alpha_2(P_1, &P_2, \sigma_n^2, \rho, \mathbf{H})=\alpha_1(P_2, P_1, \sigma_n^2, \rho, [\mathbf{h}_2, \mathbf{h}_1]),\notag\\
\beta(P_1,&P_2,\sigma_n^2,\rho,\mathbf{H}) = (\sigma_n^4 + \sigma_n^2\bar{\upsilon} + \omega)^2 \notag.
\end{align}

As observed, the denominator $\beta(\cdot)$ is always positive because it is a quadratic form. Hence, the direction of the gradient functions will only depend on the numerator terms $ \alpha_1(\cdot)$ and $\alpha_2(\cdot)$. Notice that the expression for $\alpha_2(\cdot)$ is identical to $\alpha_1(\cdot)$ but with the powers $P_1$ and $P_2$ and the columns in the channel matrix exchanged.

To obtain some insight into the problem, we analyze some particular scenarios:
\begin{itemize}
	\item For uncorrelated sources ($\rho = 0$), \eqref{eq:diff_P1} simplifies to
	\begin{align*}
	\alpha_1(P_1,&P_2,\sigma_n^2,0,\mathbf{H}) =  \\
	&- \sigma_n^4\Vert \B{h}_1 \Vert^2 - 2\sigma_n^2|\B{H}^H\B{H}|P_2 - \Vert\B{h}_2\Vert^2|\B{H}^H\B{H}|P_2^2.
	\end{align*}
	\item On the contrary, for fully correlated sources ($\rho=1$), \eqref{eq:diff_P1} simplifies to
	\begin{align*}
	\alpha_1(P_1,P_2,\sigma_n^2,1,\mathbf{H})  &= - 2\sigma_n^4\Vert \B{h}_1 \Vert^2  - 
	2\sigma_n^4|\B{h}_1^H\B{h}_2|\sqrt{\frac{P_2}{P_1}}.
	\end{align*}
	\item When considering orthogonal access channels where $\mathbf{h}_1^H\mathbf{h}_2=0$, \eqref{eq:diff_P1} simplifies to
	\begin{align*}
	\alpha_1(P_1,& P_2, \sigma_n^2, \rho, \mathbf{H}) =-\sigma_n^4(1+\rho^2)-\sigma_n^4(1+\rho^2) \Vert \B{h}_1 \Vert^2 \notag\\
	  &- 2\sigma_n^2(1-\rho^2)|\B{H}^H\B{H}|P_2 - (1-\rho^2)^2\Vert\B{h}_2\Vert^2|\B{H}^H\B{H}|P_2^2.
	\end{align*}
\end{itemize}

In the previous three scenarios, the numerator $\alpha_1(\cdot)$  is always lower than zero regardless of the channel conditions and the source correlation because $\sigma_n^2$, $\Vert \B{h}_1 \Vert^2$, $\Vert \B{h}_2 \Vert^2$, $P_1$, $P_2$, $|\B{H}^H\B{H}|$ and $|\B{h}1^H\B{h}_2|$ are all strictly positive. From \eqref{eq:KKT1} we conclude that $\lambda_1>0$, and therefore $P_1$ must be equal to $T_1$ to satisfy the condition in \eqref{eq:KKT3}. Similar reasoning applies to $P_2$. Hence, increasing simultaneously the power allocated to both users lowers the sum-\ac{MSE}, achieving a minimum when each user transmits with all its available power.

We now consider the general case. The numerators $\alpha_1(\cdot)$ and $\alpha_2(\cdot)$ are no longer necessarily negative and, for a given channel realization $\B{H}$ and source correlation $\rho$, their sign depends on the power allocated to each user, $P_1$ and $P_2 $. For convenience, we rewrite \eqref{eq:diff_P1} as follows
\begin{align}
\alpha_1(P_1,P_2,\sigma_n^2,\rho,\mathbf{H}) = A + &B\left(\Vert\B{h}_1\Vert^2P_1 - \Vert\B{h}_2\Vert^2P_2\right)\sqrt{\frac{P_2}{P_1}} \notag\\&- 2\sigma_n^4\rho |\B{h}_1^H\B{h}_2|\sqrt{\frac{P_2}{P_1}},
\label{eq:diff_P1_v2}
\end{align}
with
\begin{align}
A = & -\sigma_n^4(1+\rho^2)\Vert \B{h}_1 \Vert^2  - 2\sigma_n^2(1-\rho^2)|\B{H}^H\B{H}|P_2 \notag \\
&- (1-\rho^2)^2\Vert\B{h}_2\Vert^2|\B{H}^H\B{H}|P_2^2, \notag\\
B = & ~\sigma_n^2\rho(1-\rho^2)|\B{h}_1^H\B{h}_2|. \notag
\end{align}
As observed, the term $A$ is negative regardless of the power allocation, the channel response, the noise variance and the source correlation. The term $B$ is larger than zero but it is multiplied by $\left(\Vert\B{h}_1\Vert^2P_1 - \Vert\B{h}_2\Vert^2P_2\right)$ and by $\sqrt{\frac{P_2}{P_1}}$ which is always positive. Hence, $\Vert\B{h}_1\Vert^2P_1 > \Vert\B{h}_2\Vert^2P_2$ is a necessary condition for $\alpha_1(\cdot) > 0$. Applying the same reasoning for $\alpha_2(\cdot)$, we conclude that it is necessary that $\Vert\B{h}_2\Vert^2P_2 > \Vert\B{h}_1\Vert^2P_1$ to guarantee that $\alpha_2(\cdot)$ is also positive. Both conditions cannot be fulfilled simultaneously, and it is hence impossible to find feasible $P_1$ and $P_2$ values such that $\alpha_1(\cdot) \ge 0$ and $\alpha_2(\cdot) \ge 0$ at the same time.

Depending on the power values, we can define the three following regions
\begin{align*}
\mathcal{R}_1 &=\{(P_1,P_2) ~ | ~\alpha_1(P_1, P_2, \sigma_n^2, \rho, \mathbf{H}) \leq 0 ~\land ~ \alpha_2(P_1, P_2, \sigma_n^2, \rho, \mathbf{H})\leq 0\}, \\
\mathcal{R}_2 &= \{(P_1,P_2) ~ | ~\alpha_1(P_1, P_2, \sigma_n^2, \rho, \mathbf{H}) \leq 0 ~\land ~ \alpha_2(P_1, P_2, \sigma_n^2, \rho, \mathbf{H}) > 0\}, \\
\mathcal{R}_3 &=\{(P_1,P_2) ~ | ~\alpha_1(P_1, P_2, \sigma_n^2, \rho, \mathbf{H}) > 0 ~\land ~ \alpha_2(P_1, P_2, \sigma_n^2, \rho, \mathbf{H})\leq 0\}.
\end{align*}
According to the analysis above, the remaining region, corresponding to the power pairs such that the terms $\alpha_1(\cdot)$ and $\alpha_2(\cdot)$ are positive at the same time, is empty. Also, when $\mathcal{R}_2$ and $\mathcal{R}_3$ exist, assigning maximum available power to both users might not be optimal.

\begin{figure}
	\centering
	\includegraphics[width=0.65\columnwidth]{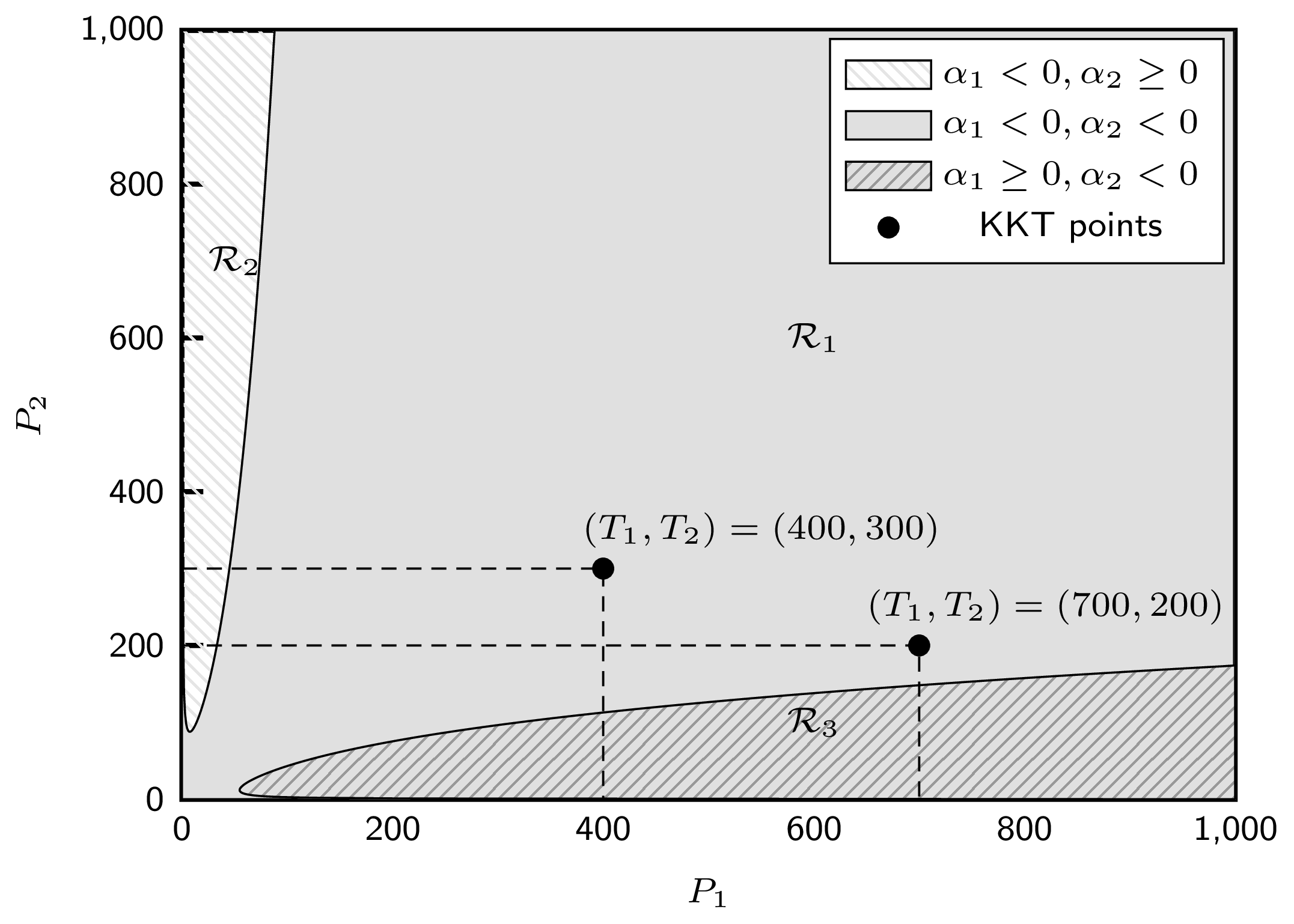}
	\includegraphics[width=0.65\columnwidth]{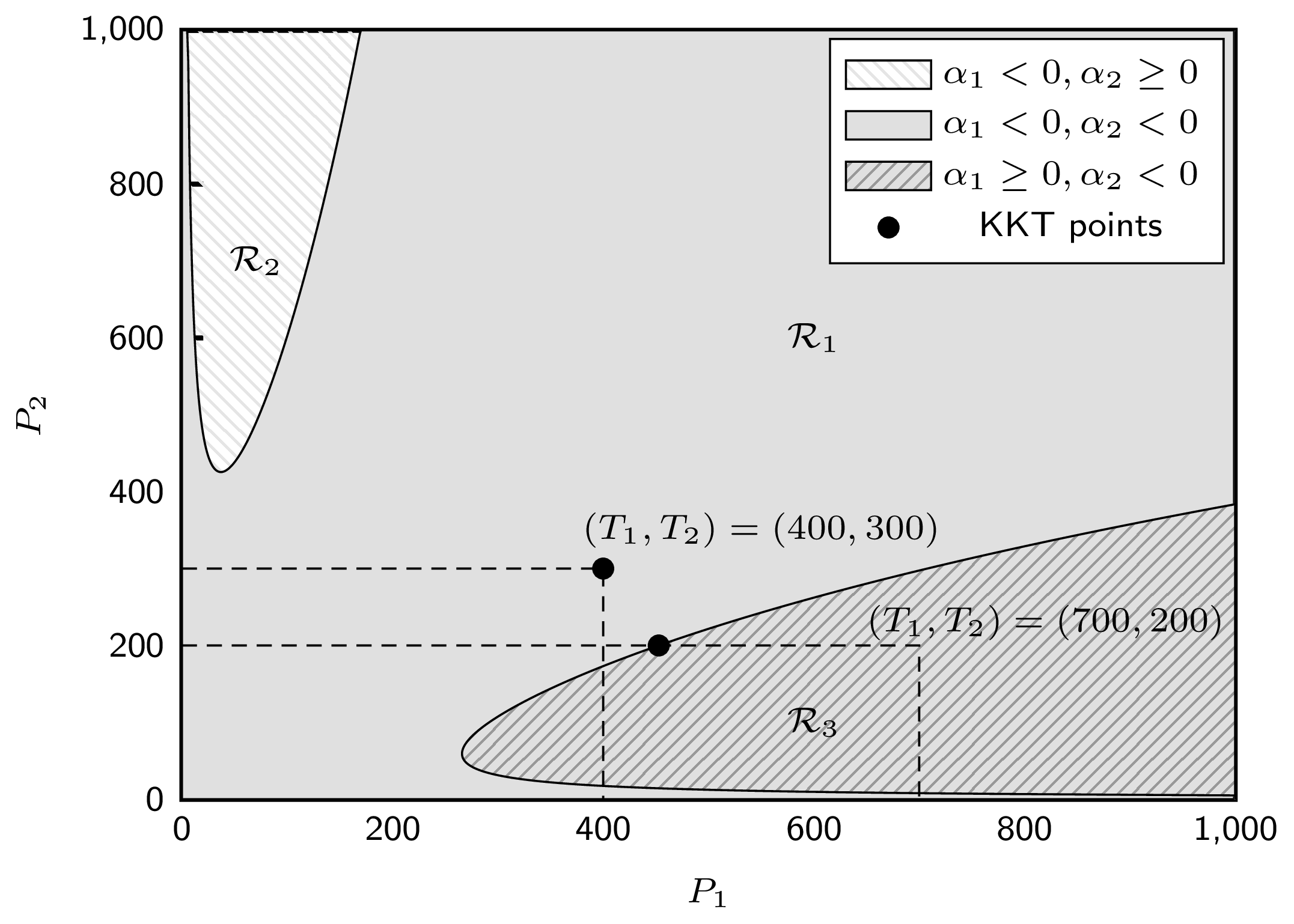}
	\caption{Example of the three feasible regions defined according to the sign of the gradient vectors for the two-user \ac{SIMO} \ac{MAC} with a correlation factor $\rho = 0.95$ (top) and $\rho = 0.99$ (bottom), for $\mathbf{h}_1=(1, 1)^T$, $\mathbf{h}_2 = (1, 0.5)^T$ and $\sigma_n^2=1$.}
	\label{fig:powerRegions}
\end{figure}

\Cref{fig:powerRegions} shows an example of the three regions described above. Each region  corresponds to a different behaviour of the gradient vectors for $P_1$ and $P_2$.
As observed, the boundary between $\mathcal{R}_1$ and $\mathcal{R}_2$ corresponds to the points where $\alpha_2(\cdot)$ is equal to zero. Similarly,  the boundary between $\mathcal{R}_1$ and $\mathcal{R}_3$ corresponds to the points where $\alpha_1(\cdot)$ is equal to zero. Therefore, the points on both boundaries can be determined by
\begin{align}
f_1(P_2, &\sigma_n^2, \rho, \mathbf{H}) = -\frac{A}{2B\Vert\B{h}_1\Vert^2\sqrt{P_2}} 
\notag\\&+\sqrt{\frac{A^2}{4B^2\Vert\B{h}_1\Vert^4P_2} + \frac{\Vert\B{h}_2\Vert^2}{\Vert\B{h}_1\Vert^2}P_2 +
	\frac{2\sigma_n^2}{(1-\rho^2)\Vert\B{h}_1\Vert^2}}, \label{eq:bound1}\\
f_2(P_1,&\sigma_n^2, \rho, \mathbf{H}) = f_1(P_1,\sigma_n^2, \rho, [\mathbf{h}_2, \mathbf{h}_1]),\label{eq:bound2}
\end{align}
where $f_1(\cdot)$ results from solving $\alpha_1(\cdot)=0$ for $\sqrt{P_1}$, and $f_2(\cdot)$ from solving $\alpha_2(\cdot)=0$ for $\sqrt{P_2}$. In this case, $f_1(\cdot)$ determines the values $\sqrt{P_1}$ where $\alpha_1(\cdot) = 0$ for a given $P_2$, while $f_2(\cdot)$ provides the values $\sqrt{P_2}$ where $\alpha_2(\cdot) = 0$ given $P_1$. Hence, with \eqref{eq:bound1} and \eqref{eq:bound2}, the regions defined above can be expressed in an alternative form as
\begin{align}
\mathcal{R}_1 &=\left\{(P_1,P_2) :\sqrt{P_1} \leq f_1(P_2, \sigma_n^2, \rho, \mathbf{H})\land \sqrt{P_2} \leq f_2(P_1, \sigma_n^2, \rho, \mathbf{H}) \right\}, \label{eq:R1} \\
\mathcal{R}_2 &= \left\{(P_1,P_2): \sqrt{P_1} \leq f_1(P_2, \sigma_n^2, \rho, \mathbf{H}) \land \sqrt{P_2} > f_2(P_1, \sigma_n^2, \rho, \mathbf{H}) \right\}, \label{eq:R2} \\
\mathcal{R}_3 &= \left\{(P_1,P_2) :\sqrt{P_1} > f_1(P_2, \sigma_n^2, \rho, \mathbf{H}) \land  \sqrt{P_2} \leq f_2(P_1, \sigma_n^2, \rho, \mathbf{H}) \right\}.
\end{align}

Since the gradient cannot be equal to zero for $P_1$ and $P_2$ simultaneously, the only feasible KKT points must lie on the boundaries defined by the power constraints. In that case, we distinguish three different situations depending on the region in which the point $(\sqrt{T_1}, \sqrt{T_2})$ falls into:	
\begin{enumerate}
	\item 	If $\left(\sqrt{T_1}, \sqrt{T_2}\right) \in \mathcal{R}_1$, the only feasible KKT point is just $(\sqrt{T_1}, \sqrt{T_2})$. On one hand, this point satisfies the KKT conditions in \eqref{eq:KKT1}, \eqref{eq:KKT2}, \eqref{eq:KKT3} and \eqref{eq:KKT4} with $\lambda_1>0$ and $\lambda_2 >0$. %
	On the other hand, from \eqref{eq:R1} and \eqref{eq:R2}, we know that points $(t, \sqrt{T_2})$ with $t<\sqrt{T_1}$ fall into $\mathcal{R}_1$ or $\mathcal{R}_2$ where $\alpha_1(\cdot)$ is always negative. In this case, condition \eqref{eq:KKT3} forces $\lambda_1=0$, and \eqref{eq:KKT1} is no longer satisfied. The same applies to points $(\sqrt{T_1}, t)$ where $t < \sqrt{T_2}$. Hence there are not more feasible KKT points on the constraints boundaries.
	
	\item If $\left( \sqrt{T_1}, \sqrt{T_2} \right) \in \mathcal{R}_2$, then it cannot be a KKT point because $\alpha_2(\cdot)>0$ in this region, and \eqref{eq:KKT2} cannot be satisfied with $\lambda_2\geq0$. Hence, the only possible KKT point can be found by lowering the power of the second user until achieving the point that makes $\alpha_2(\cdot) = 0$. Hence $\sqrt{P_2} = f_2(P_1, \sigma_n^2, \rho, \mathbf{H})$. In that case, the point $(\sqrt{T_1}, f_2(T_1, \sigma_n^2, \rho, \mathbf{H}))$ satisfies the KKT conditions with $\lambda_2=0$ and $\lambda_1>0$. 
	
	\item By symmetry, the same reasoning applies if $\left( \sqrt{T_1}, \sqrt{T_2} \right) \in \mathcal{R}_3$ getting $(f_1(T_2, \sigma_n^2, \rho, \mathbf{H}),\sqrt{T_2})$.
\end{enumerate}
It can be concluded from the previous remarks that the feasible KKT points for the optimization problem in \eqref{eq:opt_problem_2users} are $\left(\sqrt{T_1}, \sqrt{T_2}\right)$, $\left(\sqrt{T_1}, f_2\left(\sqrt{T_1}\right)\right)$ and $\left(f_1\left(\sqrt{T_2}\right), \sqrt{T_2}\right)$, leading to the power allocations stated in the Lemma.
\end{proof}
\end{lemma}

Fig. \ref{fig:powerRegions} depicts a geometric representation of the KKT points depending on the shape of each region for a given channel and two different correlations, $\rho = 0.95$ and $\rho = 0.99$. As observed, when the power constraints are set to $(T_1, T_2) = (400, 300)$, the corresponding point falls into $\mathcal{R}_1$ for both correlation factors, and hence the optimal solution allocates all the available power to both users. When the power constraints are set to $(T_1, T_2) = (700, 200)$, the corresponding point again falls into $\mathcal{R}_1$ for $\rho = 0.95$, but it falls into $\mathcal{R}_3$ for $\rho = 0.99$. Thus, increasing the source correlation causes the optimal power allocation to be $(P_1,P_2) = (452.73, 200)$. Moreover, the optimal sum-MSE for the case $(T_1, T_2) = (700, 200)$ and $\rho=0.95$ is $\xi_{\text{sum}}^o=0.027$, and it decreases to $\xi_{\text{sum}}^o=0.010$ when $\rho=0.99$, although the power allocated to the first user is lower in the latter case. A similar result would be obtained for power constraints falling into $\mathcal{R}_2$, but in this case decreasing the power allocated to the second user.

In summary, using Lemma \ref{lemma:opt_phases} and Lemma \ref{theorem:power_alloc}, the optimal precoders that minimize the sum-MSE for the two-user \ac{SIMO} \ac{MAC} are given by
\begin{itemize}
\item If $\rho=0$, $\rho=1$ or $|\B{h}^H_1\B{h}_2|=0$
\begin{align}
\B{P}^o=\left[
\begin{array}{cc}
\sqrt{T_1}& 0\\
0 & \sqrt{T_2}e^{-j\arg(\B{h}_1^H\B{h}_2)}
\end{array} 
\right].
\label{eq:precoder_2users_special}
\end{align}
\item In any other case
\begin{align}
\B{P}^o=\left[
\begin{array}{cc}
\min\left(\sqrt{T_1}, f_1\left(T_2\right)\right) & 0\\
0 & \min\left(\sqrt{T_2}, f_2\left(T_1\right)\right) e^{-j\arg(\B{h}_1^H\B{h}_2)}
\end{array} 
\right].
\label{eq:precoder_2users}
\end{align}
\end{itemize}
Notice that combining this result with the \ac{MRT} and Nu-SVD precoders described in \Cref{subsec:MIMOPrecoding} allows to analytically determine the gain factors $\gamma_1$ and $\gamma_2$ for the two-user \ac{MIMO} \ac{MAC} by simply replacing $\B{h}_k$ by the equivalent channel responses $\tilde{\B{h}}_k = \B{H}_k\B{u}_k$. 

\subsection{Sum-MSE Analysis}
In this subsection, we show some asymptotic results on the sum-MSE for the two-user MAC when optimal linear precoding is considered. We study the impact of the number of receive antennas assuming that entries of the channel matrix are i.i.d. according to a zero-mean unit-variance complex-valued Gaussian distribution, i.e., $[\B{H}]_{i,j}\sim \mathcal{N}_{\mathbb{C}}(0, 1)$:
\begin{itemize}
	\item When $N_R\rightarrow \infty$, the sum-\ac{MSE} is given by
\begin{align*}
	\bar{\xi}_{\text{sum}}(\sigma_n^2,\rho,\mathbf{H}) \approx  &\frac{2\sigma_n^4 + (1- \rho^2) \sigma_n^2N_R(T_1 + T_2)}
	{\sigma_n^4 + \sigma_n^2N_R(T_1
	+ T_2) + T_1T_2N_R^2(1-\rho^2)},
\end{align*}
the sum-MSE decreases when the number of receive antennas increases, since the denominator grows quadratically and the numerator linearly. Moreover, the impact of increasing the number of receive antennas on the performance varies with the source correlation. This will be checked experimentally in the results section.

\item When $N_R=1$, the term depending on the determinant of $\B{H}^H\B{H}$ in \eqref{eq:mse_2users} equals to zero, and we obtain the following expression for the sum-MSE
\begin{align}
	\bar{\xi}_{\text{sum}}(P_1,&P_2,\sigma_n^2,\rho,\mathbf{H}) = \notag \\&\frac{2\sigma_n^2 + (1- \rho^2) (P_1|h_1|^2 + P_2|h_2|^2)}
	{\sigma_n^2 + P_1|h_1|^2
	+ P_2|h_2|^2 + 2\sqrt{P_1P_2}\rho|h_1^Hh_2| }.
	\label{eq:twoUsersNr1}
\end{align}
As observed, for fixed $P_1$ and $P_2$, the impact of increasing the correlation factor $\rho$ is more significant in this case because it implies to simultaneously decrease the numerator and to increase the denominator, which contribute to lower the sum-MSE jointly. 
As previously shown, when $N_R\rightarrow\infty$, both numerator and denominator decrease with correlation.

\end{itemize}

\section{Simulation Results} \label{sec:results}

In this section, the performance of the precoding techniques proposed for correlated sources is evaluated for different \ac{MAC} scenarios by means of computer simulations. In these simulations, $M$ vectors of $K$ source symbols are generated from a multivariate complex-valued circularly symmetric Gaussian distribution with zero mean and covariance matrix $\B{C_s}$. Then, each user individually precodes its source symbols and transmits them over a specific realization of the fading \ac{MAC}. The channel realizations are assumed to be independent and identically distributed complex-valued circularly symmetric Gaussian random variables with zero mean and unit variance, i.e., Rayleigh fading. At the receiver, each vector of observed symbols is employed to obtain an estimate of the original symbols using the corresponding linear filter. Finally, the sum-MSE between the source and the decoded symbols for the $l$-th channel realization is estimated as
$\hat{\xi}^l_{\text{sum}} = \frac{1}{M}\sum_{m=1}^M\sum_{k=1}^K (s_{k,m} - \hat{s}_{k,m})^2.$
Each point of the performance curves is obtained after repeating the previous computer experiment over $L$ different channel realizations and averaging the resulting estimated sum-MSE distortions as
$\hat{\xi}_{\text{sum}} = \frac{1}{L}\sum_{l=1}^L \hat{\xi}^l_{\text{sum}}.$
In particular, the different experiments consist of the transmission of blocks of $M=1000$ source symbols over $L=1000$ channel realizations. The performance indicator in the figures of this section is the sum-\ac{SDR} normalized by the number of users $K$, which is defined as
\begin{equation}
	\text{SDR (dB)} = 10 \log_{10} (K/\hat{\xi}_{\text{sum}}).
\end{equation}
This performance indicator is usually plotted with respect to the \ac{SNR}. Without loss of generality, we assume $\sigma_n^2=1$ and, therefore, the \ac{SNR} per user is directly given by its power constraint: $\text{SNR}_k ~ \text{(dB)} = 10\log_{10}(T_k) ~\forall k$.   

In this section, we present the performance of the proposed linear precoding schemes in terms of the obtained \ac{SDR} for different ranges of \ac{SNR}. We also plot an upper bound for the achievable SDR under the assumption of separate source and channel coding, whose computation is described in Appendix A. Also, in some scenarios we show the performance of a transmission scheme that uses non-linear modulo functions to transform the source symbols \cite{Suarez17}, prior to being encoded with the linear strategies described in this work. %
In this case, the modulo mappings are optimized for the equivalent channel matrix resulting from the product of precoding and channel. Note that this approach can be seen as a scheme that separately performs non-linear source encoding and linear channel encoding. This strategy helps to gain some insight on the differences observed with respect to the separation bound. 

\begin{figure}
	\centering
	\includegraphics[]{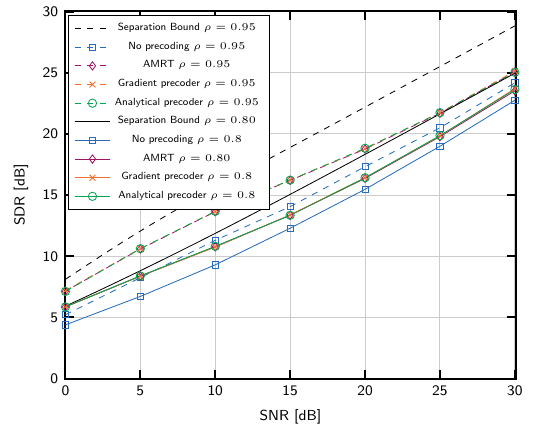}
	\caption{Performance with and without precoding for a two-user \ac{SIMO} \ac{MAC} with $N_R = 2$ receive antennas and equal power constraints. Two different correlation factors are considered: $\rho = 0.8$ and $\rho = 0.95$.}
	\label{fig:2x2}
\end{figure}

In the first computer experiment, we consider a two-user \ac{SIMO} \ac{MAC} scenario with $N_R$ = 2 receive antennas. The power constraints are identical for the two users, i.e. $\text{SNR}_1 = \text{SNR}_2 = \text{SNR}$, and $\rho$ is the correlation between the two users symbols. \Cref{fig:2x2} shows the performance curves obtained when the source symbols are transmitted with the analytical precoder given by \eqref{eq:precoder_2users} for two different correlation factors: $\rho = 0.8$ and $\rho = 0.95$. These curves are compared to those obtained with other three approaches: 1) scaling by the available powers (i.e. without precoding), 2) low-complexity \ac{AMRT} in \Cref{sec:amrt}, and 3) the projected gradient algorithm described in \Cref{sec:Gradient}. %
On one hand, the optimal linear precoder provides a performance gain for all range of SNRs with respect to no precoding, especially for low and medium SNRs. As expected, this gain is larger for higher correlation factors, since it ranges between $0.7$ dB and $1.6$ dB for $\rho = 0.8$, and between $1$ dB and $2.5$ dB for $\rho = 0.95$.
On the other hand, the AMRT and the projected gradient algorithms provide the same performance as that of the optimal precoder. Note that 
AMRT and no precoding allocate all the available power to both user. Thus the gain of the AMRT with respect to no precoding comes just from adjusting the phase of the encoder coefficients. Also, for the two-user SIMO MAC, the phase computed by AMRT agrees with the optimal one.
According to the AMRT definition in \eqref{eq:alignedPrec} and assuming, without loss of generality, that $\B{g} = \lambda\B{h}_1$, the scalar gains are $p_1 = \frac{\sqrt{T_k}}{\|\B{h}_1^H\B{g}\|}\B{h}_1^H\B{g} = \sqrt{T_1}$ and $p_2 =\frac{\sqrt{T_k}}{\|\B{h}_2^H\B{g}\|}\B{h}_2^H\B{g}= \frac{\sqrt{T_k}}{\|\B{h}_2^H\B{h}_1\|}\B{h}_2^H\B{h}_1$, and we can see that they comply with the optimality condition in \eqref{eq:rotator_precoders}. We then conclude that in this scenario with equal power constraints, the gain comes mainly from an appropriate adjusting of the encoder phases.
Finally, the gap with respect to the separation bound is especially small for low SNRs, although it gradually increases with the SNR.

\begin{figure}
	\includegraphics[scale=0.9]{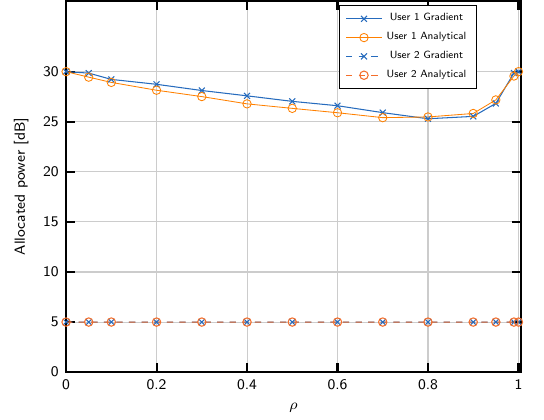}
	\includegraphics[scale=0.9]{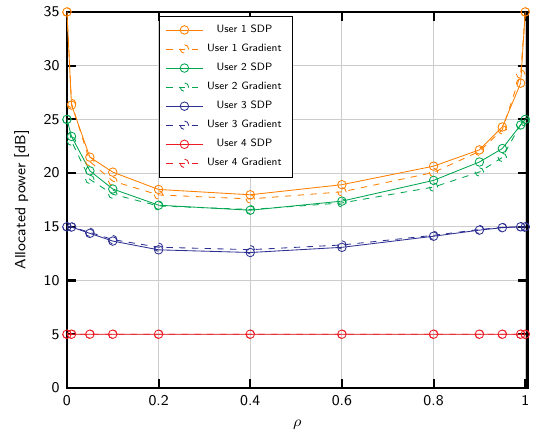}
	\caption{Average power allocated to each user after precoding the source symbols for a two-user \ac{SIMO} \ac{MAC} with $N_R = 2$ antennas when the power constraints are set such that $\text{SNR}_1=30$ dB and $\text{SNR}_2=5$ dB (top). Four-user SISO \ac{MAC} with $N_R = 1$ antennas and power constraints $\text{SNR}=[35, 25, 15, 5]$ (bottom).}
	\label{fig:powerAllocation_2x2}
\end{figure}

\Cref{fig:powerAllocation_2x2} plots the average power allocated to each user depending on the source correlation for two scenarios: 1) a two-user \ac{SIMO} \ac{MAC} with $N_R = 2$ antennas, when the individual power constraints are set such that $\text{SNR}_1=30$ dB and $\text{SNR}_2=5$ dB, 2) a four-user \ac{MAC} with $N_T=1$ and $N_R=1$ antennas, and power constraints such that the user SNRs are 35, 25, 15 and 5 dB. The optimal precoders in the first scenario are obtained, for each channel realization,  using the analytical expression in \eqref{eq:precoder_2users} and the projected gradient algorithm explained in \Cref{sec:Gradient}.  Then, the power allocated to each user is averaged over all channel realizations.
As observed, the user with the most restrictive power constraint transmits at its maximum power to lower the sum-MSE. The power allocated to the other user depends on the correlation factor. As explained in \Cref{sec:TwoUsers}, this power is the maximum available when the correlation factor is $\rho=0$ and $\rho = 1$. However, the users with the largest power budgets should not use all their available power for intermediate correlation factors.
In the second scenario, the linear precoders are obtained with the semidefinite relaxation described in \Cref{sec:SDR} and also with the gradient algorithm. The results suggest that the optimal power allocation policy for this scenario is similar to that of the previous case, and the users with more available power can benefit from the source correlation to save power for a broad range of correlation factors. This also justifies the approach proposed in \Cref{subsec:MIMOPrecoding}, which is based on the assumption that allocating all the available power to the users is not necessarily optimal when $\rho>0$.

\begin{figure}
\centering
	\includegraphics[]{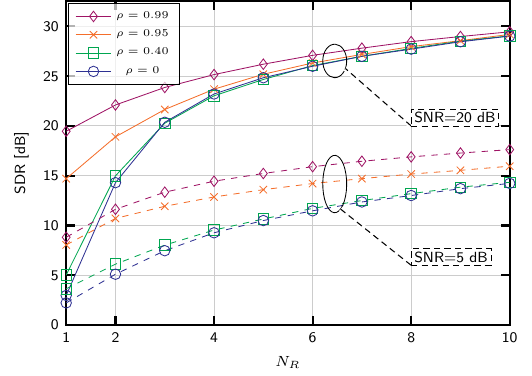}
	\caption{Performance of the optimal linear precoder depending on the number of receive antennas for the two-user \ac{SIMO} \ac{MAC}, for SNR=5 dB and SNR=20 dB, and different correlation factors.}
	\label{fig:Fig2-1xNR_SNR20}
\end{figure}

\Cref{fig:Fig2-1xNR_SNR20} shows the performance of the optimal linear precoder for the two-user SIMO MAC with a different number of receive antennas, for SNR=5 dB and SNR=20 dB. Separation bounds are omitted for clarity.  An interesting result is observed when the source symbols are highly correlated since high SDR values are obtained even for a small number of receive antennas. As seen in \eqref{eq:twoUsersNr1}, the phase of the user precoders is designed to exploit the source correlation by aligning the user channels, which contributes to lower the sum-MSE, especially for $N_R=1$ and high correlation factors. The impact of increasing the receive antennas on the SDR is more remarkable for uncorrelated sources, although the performance gain diminishes for a large number of receive antennas regardless of the source correlation. Moreover, the gain due to the correlation with a large number of receive antennas vanishes for high SNRs, while a small gain is still observed for low SNRs. This agrees with the asymptotic analysis shown in \Cref{sec:precoder_sumMSE}. 

\begin{figure}
	\centering
	\includegraphics[]{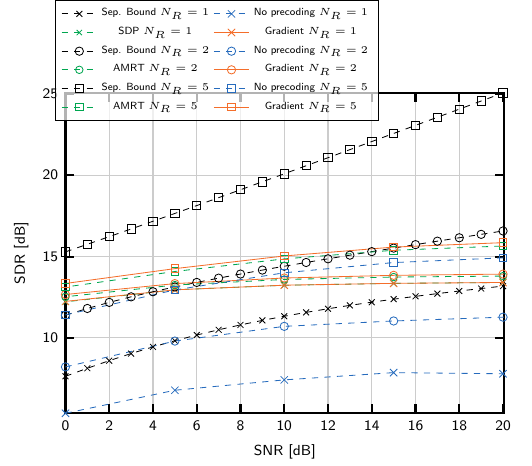}
	\caption{Performance with different precoding strategies for a \ac{SIMO} \ac{MAC} with and without precoding depending on the number of receive antennas $N_R$ with $K=10$ users, $\rho = 0.95$ and equal power constraints.}
	\label{fig:comNR}
\end{figure}

We now consider a \ac{SIMO} \ac{MAC} scenario with $K=10$ users. In such a case, the optimal sum-MSE precoders cannot be obtained via analytical expressions. Hence, we show results for the projected gradient algorithm, the \ac{SDP} in \Cref{sec:SDR}, and the aligned \ac{MRT} algorithm in \Cref{sec:amrt}. \Cref{fig:comNR} shows the performance curves with and without precoding for three different number of receive antennas: $N_R = 1$, $2$ and $5$. The power constraints are equal for all the users and the source correlation is $\rho_{i,j}=\rho = 0.95, \forall i\neq j$. The corresponding bounds based on source-channel separation are also shown. As observed, the performance achieved by the \ac{AMRT} and the \ac{SDP} is close to the one provided by the projected gradient algorithm. The gain provided by these precoders is more significant as the number of receive antennas is smaller, and the extreme case occurs when $N_R=1$ where the precoding gain is substantial (about $8$ dB). Increasing the number of receive antennas improves the overall performance in general, but the gain provided by the different precoders is not so significant with respect to the uncoded case. These results agree with the behavior observed for two users and confirm the suitability of linearly encoding the source symbols for scenarios with a smaller number of receive antennas than users and correlated sources.
It is also interesting to remark that for low SNR values, the precoding schemes improve the bound based on source-channel separation, showing that some low computational complexity strategies such as \ac{AMRT} can beat separate coding. 

\begin{figure}
	\centering
	\includegraphics[]{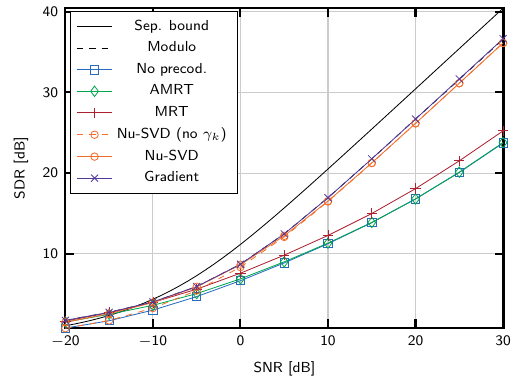}
	\caption{Performance with different precoding strategies for a \ac{MIMO} \ac{MAC} with $K = 10$, $N_T=2$, $N_R = 10$, $\rho = 0.40$ and equal power constraints.}
	\label{fig:FigSdr10-2x10EqualCorrEqualPower0.80}
\end{figure}

\begin{figure}
	\centering
	\includegraphics[]{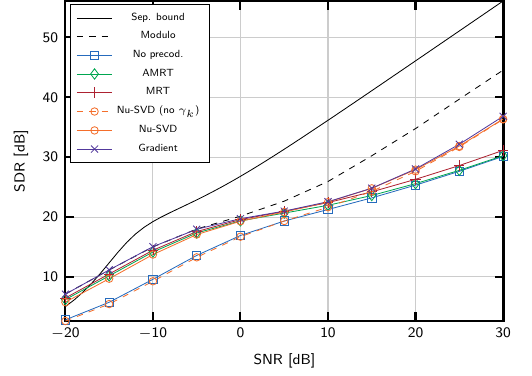}
	\caption{Performance with different precoding strategies for a \ac{MIMO} \ac{MAC} with $K = 10$, $N_T=2$, $N_R = 10$, $\rho = 0.99$ and equal individual constraints.}
	\label{fig:FigSdr10-2x10EqualCorrEqualPower0.99}
\end{figure}
	
In the ensuing experiments, we consider a \ac{MIMO} \ac{MAC} scenario with $K=10$ users with $N_T=2$ antennas per user and equal power constraints, and a receiver with $N_R=10$ antennas. The correlation factor is $\rho_{{i,j}}=\rho=0.40 ~ \forall i,j, ~ i\neq j$. Figure \ref{fig:FigSdr10-2x10EqualCorrEqualPower0.80} plots the performance curves when the linear precoder is obtained with the following strategies: 1) projected gradient algorithm, 2) \ac{AMRT}, 3) optimized \ac{MRT}, 4) optimized Nu-SVD and 5) Regular Nu-SVD. Recall that for the optimized version of MRT and Nu-SVD, the factor gains $\gamma_k$ are adjusted as explained in Section \ref{subsec:MIMOPrecoding} to exploit the source correlation. At the receiver, the linear \ac{MMSE} filter, given by \eqref{eq:MMSEestim}, is employed for all the precoding methods. On one hand, the precoder obtained with the projected gradient algorithm provides the best performance for all the considered SNR values. The optimized Nu-SVD scheme closely approaches this performance for all SNR values, providing certain gain (about 1-2 dB) with respect to the regular Nu-SVD up to SNR=0 dB, due to the adjusting of the factor gains. 
Optimized MRT also obtains the same performance as the gradient precoder and Nu-SVD for SNR values below 0 dB, but the gap with the other strategies increases with the SNR. Low-complexity AMRT behaves in a similar way to the previous strategies at low SNR regimes, but it converges to the uncoded scheme for SNR $\ge 0$ dB since the correlation between the source symbols is low. 
On the other hand, projected gradient and Nu-SVD performance stay close to the separation bound, due to the low correlation factor, while the modulo-based encoding of the source symbols does not contribute to improving the performance. Also, the separation bound shown for $K>2$ users does not take into account individual constraints and the bound calculation is only based on equating the sum-distortion and the sum-capacity. For this reason, the separation bound is not necessarily tight and an additional gap between the performance curves and the bound can be observed. In spite of that, linear precoder exceeds the performance of any separation-based scheme in the low SNR regime.

Fig. \ref{fig:FigSdr10-2x10EqualCorrEqualPower0.99} plots the performance curves for the previous scenario but for higher correlation $\rho=0.99$. Again, the best performance is achieved with the precoders obtained with the projected gradient algorithm, and the optimized Nu-SVD precoding scheme also performs closely for all SNR values. In this case, adjusting the $\gamma_k$ factors significantly improves the SDR obtained with the regular Nu-SVD (about $5$ dB for low SNRs) due to the higher source correlation.
\ac{MRT} also takes advantage of the high correlation and performs close to those schemes for SNRs below $10$ dB, lowering the gap for high SNRs with respect to $\rho=0.4$. Similar behavior is observed for \ac{AMRT}, although for large SNR values its performance is again closer to no precoding. In this scenario, the gap with  respect to the separation bound is larger than in the previous experiment. %
It is interesting to observe that, unlike for $\rho=0.4$, the combination of modulo mappings and linear precoding provides significant improvement with respect to directly precoding the source information, especially for high SNRs. This combined strategy allows to reduce the gap to the separation bound for SNR values over 0 dB, and this gain can be attributed to the non-linear nature of the modulo mapping. Different works \cite{lapidoth10, floor15} have shown that the linear precoder of source symbols is optimal for SNRs lower than a particular value. However, this SNR threshold diminishes as the source correlation increases and, in this case, non-linear mappings are required to approach the theoretical limits.

\section{Conclusion} \label{sec:conclusions}

Linear distributed encoding for the transmission of correlated information over fading \ac{MIMO} \acp{MAC} has been considered. Precoders are designed to minimize the sum-MSE considering the source correlation which results in a non-convex optimization problem. We have proposed different strategies to obtain the linear precoders at each user in the general MIMO MAC scenario. The convergence of these solutions to the optimal minimum is not guaranteed, but they provide good performance in different situations and various computational complexity levels.
In the more restrictive case of two-user \ac{SIMO} \ac{MAC}, we have derived a closed-form expression for the optimal precoder. A significant performance gain is achieved by adjusting the precoder phase difference between the two nodes and with an adequate power allocation. Regarding power allocation, an interesting fact has been observed: the nodes must transmit with all the available power when sources are either uncorrelated and fully correlated, while for other correlation levels lower power can be used. Computer simulations show a significant performance improvement regarding \ac{SDR} with respect to the uncoded case, especially for low SNRs and numbers of transmitting nodes larger than the number of receive antennas, where it is possible to achieve large SDR values by exploiting the source correlation. Hence, the proposed strategies are useful in those communication scenarios such as \acp{WSN}, where the simplicity and the power consumption of the nodes is a critical requirement.
\section*{Acknowledgment}

This work has been funded by Office of Naval Research Global of United States (N62909-15-1-2014), the Xunta de Galicia (ED431C 2016-045, ED341D R2016/012, ED431G/01), the Agencia Estatal de Investigación of Spain (TEC2015-69648-REDC, TEC2016-75067-C4-1-R) and ERDF funds of the EU (AEI/FEDER, UE).

\appendices

\section{Upper Bound based on Source-Channel Separation}\label{app:opta_sep}

The separation bound is calculated by equating the source rate-distortion region and the capacity region of the fading MAC.
Given a set of distortion targets $\B{d} = [ D_1,D_2,...,D_N ]^T$, the rate distortion region $\mathcal{R}(\B{d})$ is the set of rate distortion functions corresponding to the individual sources, i.e. $R_k^D(D_i), k = 1, . . . , K$, and by the sum-rate distortion function $R_{\text{sum}}^D (\mathcal{D})$. 
For the case of distributed encoding of  bivariate Gaussian sources under the MSE distortion, this region is defined in \cite{Oohama97,Wagner08}. For scenarios with more than two source symbols, \cite{Wang2010} provides the following lower bound for the sum-rate %
\begin{align}
R_{\text{sum}}^D(\B{d}) = \min_{\B{D}_{\B{s}}:d_{ii}\le [\B{d}]_i}~~\log\left(\frac{\vert \B{C}_\B{s}\vert}{\vert \B{D}_\B{s} \vert}\right),
\label{eq:rateD_distributed}
\end{align} 
where $\B{D}_{\B{s}}= (\B{C}_s^{-1} + \B{B})^{-1}$, for some diagonal matrix $\B{B}$, and $\B{C}_\B{s}$ is the source covariance matrix. The whole rate-distortion region is defined by \eqref{eq:rateD_distributed} applied to all user subsets, but its computation is unaffordable for large number of users.
In this case, the separation bound can be approximated by equating only the sum-distortion function to the sum-capacity of the MAC. In general, this bound will be optimistic since the individual rate constraints are not necessarily satisfied. 

The sum-capacity of a MIMO MAC with non-cooperative users is  
\begin{align}
R_{\text{{sum}}}^C(\mathbf{H}) = \log \left| \mathbf{I} + \frac1{\sigma_n^2}\sum_{k=1}^K\mathbf{H}_k\B{Q}_k\mathbf{H}_k^H \right|,
\label{eq:cap_separation_simo}
\end{align}
where the covariance matrices $\B{Q}_k$ can be computed with the iterative waterfilling algorithm \cite{Yu04}, with the constraints $\operatorname{tr}(\B{Q}_k) < T_k, \forall k$.

In general, the separation bound can be defined in terms of intersections between the rate distortion and capacity regions. For two users,  we state the following minimization problem
\begin{align}
D_{\text{sum}}(\B{H}) = \min_{D_1,D_2:\mathcal{C}(\B{H}) \cap \mathcal{R}(D_1,D_2) \neq \emptyset}~~D_1+D_2,
\end{align}
where $\mathcal{C}(\B{H})$
represents the capacity region
while $\mathcal{R}(D_1,D_2) = \mathcal{R}_1^D(D_1) \cap \mathcal{R}_2^D(D_2) \cap \mathcal{R}_{\text{sum}}^D(D_1,D_2)$ is the corresponding rate distortion region for the bivariate Gaussian distribution. For $K>2$,  we will only consider the constraints on the sum rates for all users, due to the computational complexity of searching for the best intersection between regions in such case. Hence, the separation bound is calculated by equating \eqref{eq:rateD_distributed} and \eqref{eq:cap_separation_simo}, i.e.,
\begin{align}
D_{\text{sum}}(\B{H}) = \min_{\B{d}:R^D_{\text{sum}}(\B{d})=R_{\text{sum}}^C(\B{H})}~~\B{1}^T\B{d},
\label{eq:upper_bound_sep}
\end{align}
and then averaging the obtained distortions for each channel realization to determine the optimum SDR, i.e.  $\text{SDR}_{\text{opt}} = \mathbb{E}_{\B{H}}[1/D_{\text{sum}}(\B{H})].$

\bibliographystyle{IEEEtranTCOM}
\bibliography{references}

\end{document}